\documentclass[review,3p,times,authoryear]{elsarticle_cosenior}

\usepackage{amssymb}
\usepackage{amsmath}

\usepackage{lineno}

\usepackage{soul}
\usepackage{url}
\usepackage{ulem}
\usepackage[table,xcdraw]{xcolor}
\usepackage{hyperref}

\hypersetup{
    colorlinks=true, 
    urlbordercolor=blue,
    pdfborderstyle={/S/U/W 1} 
}

\makeatletter
\newcommand{\semismall}{\@setfontsize{\semismall}{9.5}{11.5}}
\makeatother

\journal{\textup{Published in} \href{https://doi.org/10.1016/j.jad.2025.120416}{\textit{Journal of Affective Disorders} \textup{(2026)}}}

\usepackage{pdfpages} 
\usepackage{pgffor} 

\makeatletter
\AtBeginDocument{\let\LS@rot\@undefined}
\makeatother

\def\supplementfilename{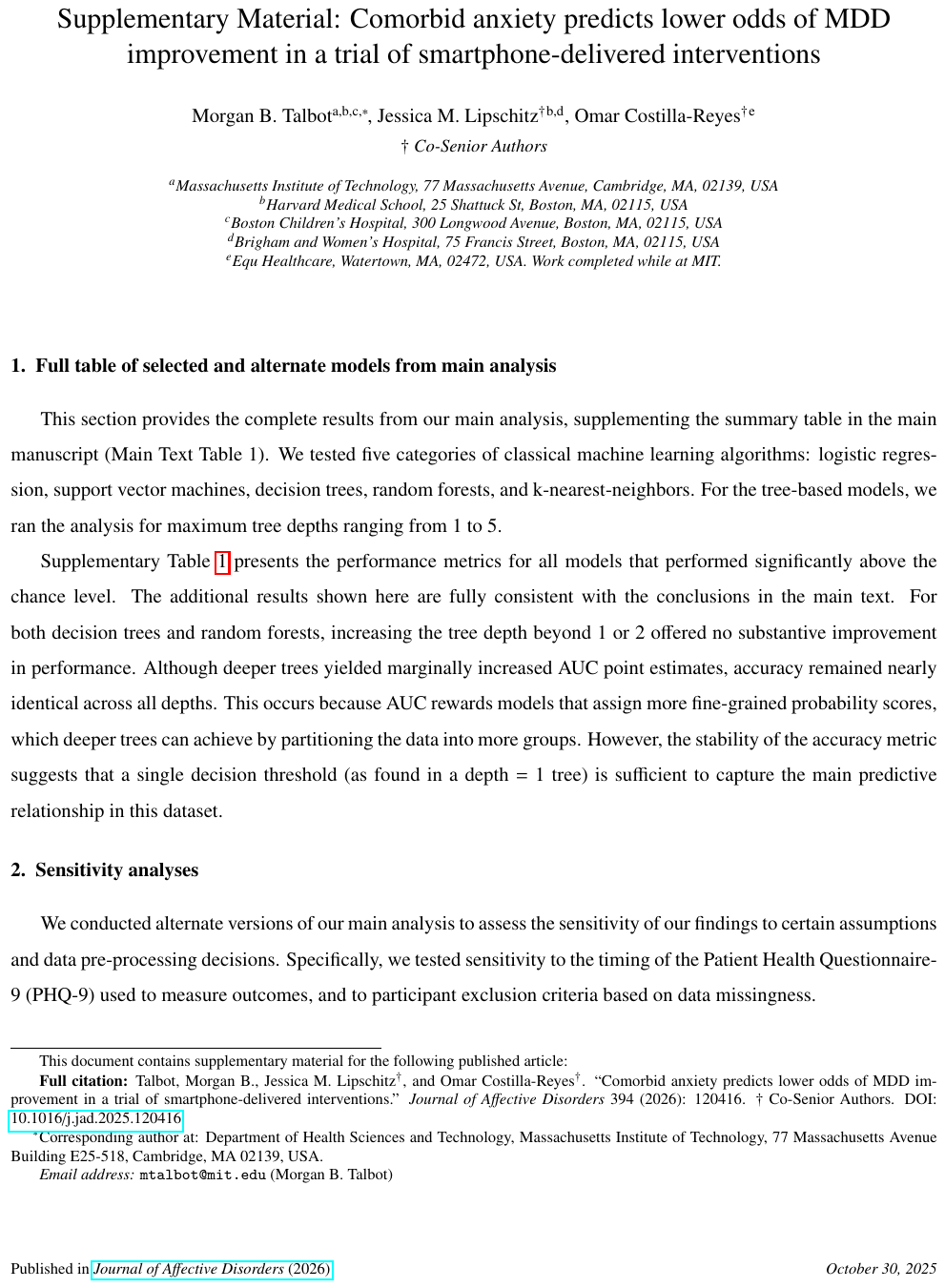}

\pdfximage{\supplementfilename}
\def\numbersupplementpages{\the\pdflastximagepages}

\newif\ifarXiv
\arXivtrue 

\begin{document}

\begin{frontmatter}



\title{Comorbid anxiety predicts lower odds of MDD improvement in a trial of smartphone-delivered interventions}

\author{Morgan B. Talbot\corref{cor1}\fnref{mit,hms,bch}}
\ead{mtalbot@mit.edu}

\author{Jessica M. Lipschitz$^\dag$\fnref{hms,bwh}}

\author{Omar Costilla-Reyes$^\dag$\fnref{equ}}

\coseniorauthors{$\dag$ \textit{Co-Senior Authors}}

\nonumnote{\textbf{Full citation:} Talbot, Morgan B., Jessica M. Lipschitz$^\dag$, and Omar Costilla-Reyes$^\dag$. ``Comorbid anxiety predicts lower odds of MDD improvement in a trial of smartphone-delivered interventions.'' \textit{Journal of Affective Disorders} 394 (2026): 120416. $\dag$ Co-Senior Authors. DOI: \href{https://doi.org/10.1016/j.jad.2025.120416}{10.1016/j.jad.2025.120416}}

\nonumnote{\textbf{Source code for data analysis:} \url{https://github.com/MorganBDT/brighten-mdd-outcome-predict}}

\cortext[cor1]{Corresponding author at: Department of Health Sciences and Technology, Massachusetts Institute of Technology, 77 Massachusetts Avenue Building E25-518,  Cambridge, MA 02139, USA.}

\affiliation[mit]{organization={Massachusetts Institute of Technology},
            addressline={77 Massachusetts Avenue},
            city={Cambridge},
            state={MA},
            postcode={02139},
            country={USA}}

\affiliation[hms]{organization={Harvard Medical School},
            addressline={25 Shattuck St},
            city={Boston},
            state={MA},
            postcode={02115},
            country={USA}}

\affiliation[bch]{organization={Boston Children's Hospital},
            addressline={300 Longwood Avenue},
            city={Boston},
            state={MA},
            postcode={02115},
            country={USA}}

\affiliation[bwh]{organization={Brigham and Women's Hospital},
            addressline={75 Francis Street},
            city={Boston},
            state={MA},
            postcode={02115},
            country={USA}}

\affiliation[equ]{organization={Equ Healthcare},
            city={Watertown},
            state={MA},
            postcode={02472},
            country={USA. Work completed while at MIT.}}


\begin{abstract}
Comorbid anxiety disorders are common among patients with major depressive disorder (MDD), but their impact on outcomes of digital and smartphone-delivered interventions is not well understood. This study is a secondary analysis of a randomized controlled effectiveness trial (n=638) that assessed three smartphone-delivered interventions: Project EVO (a cognitive training app), iPST (a problem-solving therapy app), and Health Tips (an active control). We applied classical machine learning models (logistic regression, support vector machines, decision trees, random forests, and k-nearest-neighbors) to identify baseline predictors of MDD improvement at 4 weeks after trial enrollment. Our analysis produced a decision tree model indicating that a baseline GAD-7 questionnaire score of 11 or higher, a threshold consistent with at least moderate anxiety, strongly predicts lower odds of MDD improvement in this trial. Our exploratory findings suggest that depressed individuals with comorbid anxiety have reduced odds of substantial improvement in the context of smartphone-delivered interventions, as the association was observed across all three intervention groups. Our work highlights a methodology that can identify interpretable clinical thresholds, which, if validated, could predict symptom trajectories and inform treatment selection and intensity. 
\end{abstract}




\end{frontmatter}



\vspace{-3mm}
\section{Introduction}
\label{sec:intro}
\vspace{-1mm}

Major depressive disorder (MDD) affects roughly 322 million people, and is a leading cause of disability with large impacts on quality of life \citep{moreno2021global}. Less than 20\% of people with MDD receive minimally adequate treatment \citep{thornicroft2017undertreatment}. 50\% of patients who receive MDD treatment experience minimal or no improvement, with a subset of these not responding to multiple treatment attempts \citep{gaynes2020defining}. Digital and smartphone-delivered psychotherapy interventions are one promising avenue to increase access to evidence-based MDD treatments \citep{linardon2024current}. Knowledge of the factors that predict which patients are likely to improve could inform personalized care, such as identifying individuals who may require additional support. However, studies attempting to identify these predictors have yielded inconsistent results \citep{sextl2024systematic}. 

In this study, we investigated predictors of clinical MDD improvement by conducting a secondary analysis of a large, publicly available clinical trial dataset \citep{arean2016use}. The Brighten MDD trial was an online, fully remote effectiveness trial with a four-week primary intervention period, in which participants were randomized to one of three conditions: Project EVO, a serious game designed to bolster cognitive skills related to MDD; iPST, an app based on problem-solving therapy for MDD; and an information control app called Health Tips, which suggested strategies to improve health and serves as an active control. Consistent with an effectiveness trial framework, some participants endorsed simultaneously receiving other treatments (e.g., medication, seeing a therapist or psychiatrist) while in the trial. The original clinical trial analysis found that for participants with moderate MDD, the active apps resulted in higher remission rates compared to the control intervention at the 12-week follow up \citep{arean2016use}. Although the original study compares effectiveness across the three groups, the participant-level factors related to the likelihood of MDD improvement have not been explored to the best of our knowledge. We applied interpretable machine learning techniques, coupled with a forward feature selection approach, to identify variables measured at baseline that predict greater or lesser odds of clinical improvement during the treatment period.

\vspace{-3mm}
\section{Methods}
\label{sec:methods}
\vspace{-1mm}

\subsection{Original clinical trial}

This study is an independent secondary analysis of open-access data from the Brighten study, a randomized controlled effectiveness trial. The original study evaluated the effectiveness of two smartphone-delivered interventions for depression—Project EVO (a cognitive training app) and iPST (a problem-solving therapy app)—against an active control intervention, Health Tips \citep{arean2016use}. Consolidated Standards of Reporting Trials (CONSORT) diagrams detailing participant enrollment and flow are available in the original publications, which also provide comprehensive details about the interventions and trial design \citep{anguera2016conducting, arean2016use}. The primary analysis of the original trial found that, among participants with moderate MDD at baseline, both the Project EVO and iPST interventions resulted in higher remission rates at the 12-week follow-up compared to the Health Tips active control \citep{arean2016use}.

\vspace{-1mm}
\subsection{Models and variables}

Our study predicted a binary MDD improvement outcome, measured 4 weeks after trial enrollment in alignment with the main intervention period of the original study \cite{arean2016use}. This outcome was defined using established criteria for MDD treatment response: a Patient Health Questionnaire-9 (PHQ-9) score of both $< 10$ and reduced by $\geq 50\%$ relative to baseline \citep{kroenke2001phq}. We chose a machine learning framework for this analysis because of its ability to systematically identify predictive patterns using both linear and non-linear models, an approach well-suited to contexts where prior evidence for predictors is inconsistent. Our methodology was designed to compare multiple algorithm types, rigorously account for missing data and sampling variability, and select a parsimonious set of predictors to generate robust and interpretable findings. We selected five commonly used algorithms to represent distinct modeling approaches: logistic regression and support vector machines as standard linear classifiers; decision trees and random forests to capture non-linear relationships and interactions; and k-nearest-neighbors as a non-parametric, instance-based method. We were specifically interested in decision trees for their ease of interpretability in clinical settings \citep{banerjee2019tree}. We considered the following variables in the dataset as ``features'' that the models could use for prediction ($> 0\%$ percentages of missing data shown in brackets after each variable name):

\begin{itemize}
    \vspace{-2mm}
    \item Demographics:
    \vspace{-2mm}
    \begin{itemize}
        \item Age
        \item Gender (collected as binary male/female)
        \item Race/ethnicity (categories ``African-American/Black,'' ``Asian,'' ``Hispanic/Latino,'' ``Multiracial/other,'' and ``Non-Hispanic White'')
        \item Working/employment (binary yes/no)
        \item Marital status (binarized to married/partnered or not)
        \item Education (binarized to education beyond high school or not)
        \item Satisfaction with level of income (binarized to responses of ``can't make ends meet'' vs. responses indicating higher satisfaction) [53\% missing]
    \end{itemize}
    \vspace{-2mm}
    \item Questionnaire scores:
    \vspace{-2mm}
    \begin{itemize}
        \item PHQ-9 at baseline \citep{kroenke2001phq} [0\% missing]
        \item Generalized Anxiety Disorder-7 (GAD-7), used as a global measure of anxiety symptoms \citep{spitzer2006brief, kroenke2007anxiety, beard2014beyond} [47\% missing]
        \item Sheehan Disability Scale (SDS) \citep{sheehan1983anxiety} [47\% missing]
        \item AUDIT alcohol consumption questions (AUDIT-C) \citep{bush1998audit} [47\% missing]
    \end{itemize}
    \vspace{-2mm}
    \item Treatment group (binary feature for each)
    \vspace{-2mm}
    \begin{itemize}
        \item Project EVO
        \item iPST
        \item Health Tips (active control)
    \end{itemize}
\end{itemize}

The set of predictor variables was chosen to include all available baseline demographic and clinical questionnaire data for an exploratory analysis. Marital status, education, and satisfaction with level of income were binarized. For race/ethnicity, response options ``Native Hawaiian/other Pacific Islander,'' ``American Indian/Alaskan Native,'' and ``More than one'' were combined to form the category ``Multiracial/other.'' These modifications were designed to prevent issues arising from categories with few participants. To maintain a parsimonious model and avoid multicollinearity from redundant predictors, we selected one of the two available income-related variables for inclusion. We chose ``satisfaction with level of income,'' hypothesizing that this measure of subjective financial distress may have a more direct relationship with mental health outcomes than absolute income brackets. We also considered using responses from the IMPACT mania and psychosis screening questionnaire \citep{unutzer2002collaborative,arean2016use} as predictors. We decided to exclude participants who endorsed a history consistent with bipolar disorder. Only 2 of the remaining participants reported any history of psychosis, precluding a meaningful analysis of psychosis history as a predictor of MDD outcomes. The most detailed source of information regarding the variables collected during the original trial is the \href{https://www.synapse.org/Synapse:syn10848316}{Brighten Study Public Researcher Portal}  \citep{BrightenStudyPortal}.

\vspace{-1mm}
\subsection{Data preparation}

Data from the Brighten Version 1 study \citep{anguera2016conducting, arean2016use} were downloaded from the \href{https://www.synapse.org/Synapse:syn10848316}{Brighten Study Public Researcher Portal}  \citep{BrightenStudyPortal}. We used all available records in the Brighten Version 1 dataset \citep{anguera2016conducting}. While demographics and baseline PHQ-9 scores were collected upon enrollment, several other questionnaire scores were collected in the days following enrollment, notably GAD-7, SDS, AUDIT-C, and mania and psychosis history. For these questionnaires, we used the earliest available response from each participant as the ``baseline'' measurement. We did not consider any such responses that were made more than 5 days after enrollment. We excluded participants with a baseline PHQ-9 score below 10. To focus our analysis on unipolar depression, we also excluded participants who endorsed a history of (i) lithium prescription, (ii) prescription of medication for mania symptoms, or (iii) diagnosis of bipolar disorder on the IMPACT questionnaire. These criteria left us with 638 participants. 

The PHQ-9 outcome data required for our analysis had substantial missingness, with 52\%, 58\%, 59\%, and 60\% missing at weeks 1-4 post-enrollment, respectively. Moreover, 41\% of the included participants had only demographic data and baseline PHQ-9 (which were collected together), without any questionnaire scores for GAD-7, SDS, IMPACT, AUDIT-C, and all post-baseline PHQ-9. We checked for differences in demographics and baseline PHQ-9 scores between this group and participants who had additional data, and did not find statistically significant evidence supporting any differences ($p > 0.05$, t-tests for age and PHQ-9, $\chi^2$ tests for gender, race/ethnicity, employment, marital status, and education). We used a random forest-based multiple imputation strategy with predictive mean matching to handle missing data (baseline variables and follow-up PHQ-9 at weeks 1-4), implemented with the miceRanger package in R \citep{wilson2020miceranger}. We produced 100 imputed versions of the dataset. On average across imputations, 41\% of the participants met the definition for MDD improvement. Before training each machine learning model, all non-categorical features in the dataset were rescaled to have a mean of 0 and a standard deviation of 1.

\vspace{-1mm}
\subsection{Feature selection and model fitting}

To minimize overfitting due to the high number of features in the dataset, we used a forward selection procedure to identify a minimal set of input features for each machine learning model. AUC estimates were first obtained for univariate models on each feature, and the feature that resulted in the highest AUC was selected. Then, all possible bivariate models including the first chosen feature were tested, and features were added one by one in this fashion until adding another feature did not significantly increase AUC. Each AUC estimate was obtained using a Monte Carlo cross-validation procedure with 10,000 iterations. For each iteration, we first randomly selected 1 of the 100 imputed versions of the dataset, and then randomly sampled 80\% of the participants to form a training set, leaving the remaining 20\% as a validation set. We took the 10,000 difference values between estimates from the current candidate model and those of the previous best model (or AUC = 0.5 for the first variable). We computed both the mean $\Delta$AUC value and its 95\% confidence interval from this empirical distribution, which accounts for uncertainty due to both missingness and recruitment sampling. Statistical significance of the model's improvement due to the added variable was assessed by calculating the probability $p = P(\Delta \text{AUC} \leq 0)$ from the empirically sampled distribution. The Benjamini-Hochberg procedure \citep{benjamini1995controlling} was applied to control the false discovery rate at 0.05 across all comparisons for a given model type. To balance the goals of maximizing prediction accuracy and limiting model complexity, we also required an AUC improvement of $\geq 0.02$ for each new variable \citep{greenberg2024predicting}. For decision trees and random forests, we ran variable selection for different maximum tree depths (maximum number of binary decisions allowed per tree, ranging here from 1 to 5 inclusive), keeping the tree depth that produced the highest AUC estimate overall. While the forward selection process was guided exclusively by AUC, we also calculated classification accuracy for the final models to aid in their interpretation and comparison. All models were implemented using the scikit-learn Python library with default hyperparameters \citep{pedregosa2011scikit}.

\vspace{-1mm}
\subsection{Sensitivity Analyses}
\label{sec:sensitivity_analyses}

We conducted two alternate versions of our analysis to assess the sensitivity of our findings to certain methodological assumptions. 

\begin{enumerate}
    \item Instead of predicting MDD outcomes 4 weeks post-enrollment (corresponding to the main intervention period in the original trial \citep{arean2016use}), we predicted outcomes at 12 weeks. 
    \item Of the 638 participants who met our screening criteria (baseline PHQ-9 $\geq 10$, no reported history of bipolar disorder), 279 (41\%) had only baseline PHQ-9 and demographic information recorded, with none of GAD-7, SDS, AUDIT-C, mania/psychosis history, or any post-baseline PHQ-9. While our primary analysis includes these participants and imputes their data in accordance with intention-to-treat principles, we exclude them in an alternative version of the analysis ($n=359$). Notably, missing data from these participants accounts for a large proportion of the overall missingness in the dataset: more information can be found in the supplementary material. 
\end{enumerate}

\vspace{-3mm}
\section{Results}
\label{sec:results}
\vspace{-1mm}

\begin{table}[t]
\semismall
\begin{tabular}{lllllll}
\textbf{Model Type}     & \textbf{Interp?}        & \textbf{AUC}                             & \textbf{Accuracy}                        & \textbf{Depth}           & \textbf{Pred.}           & \textbf{Coefficient}                        \\ \hline
Logistic Regression     & Yes                     & 0.74 (0.65, 0.82)                        & 0.69 (0.61, 0.77)                        & -                        & GAD-7                        & -0.96 (-1.13, -0.80)                        \\
{\color[HTML]{9B9B9B} } & {\color[HTML]{9B9B9B} } & {\color[HTML]{9B9B9B} 0.71 (0.62, 0.79)} & {\color[HTML]{9B9B9B} 0.66 (0.58, 0.74)} & {\color[HTML]{9B9B9B} -} & {\color[HTML]{9B9B9B} SDS}   & {\color[HTML]{9B9B9B} -0.75 (-0.93, -0.57)} \\ \hline
Support Vector Machine  & Yes                     & 0.74 (0.65, 0.82)                        & 0.70 (0.62, 0.78)                        & -                        & GAD-7                        & -1.01 (-1.15, -0.88)                        \\
{\color[HTML]{9B9B9B} } & {\color[HTML]{9B9B9B} } & {\color[HTML]{9B9B9B} 0.70 (0.61, 0.79)} & {\color[HTML]{9B9B9B} 0.66 (0.57, 0.74)} & {\color[HTML]{9B9B9B} -} & {\color[HTML]{9B9B9B} SDS}   & {\color[HTML]{9B9B9B} -0.81 (-0.98, -0.58)} \\ \hline
Random Forest           & No                      & 0.74 (0.65, 0.82)                        & 0.70 (0.62, 0.78)                        & 2                        & GAD-7                        & -                                           \\
{\color[HTML]{9B9B9B} } & {\color[HTML]{9B9B9B} } & {\color[HTML]{9B9B9B} 0.70 (0.61, 0.79)} & {\color[HTML]{9B9B9B} 0.66 (0.58, 0.74)} & {\color[HTML]{9B9B9B} 2} & {\color[HTML]{9B9B9B} SDS}   & {\color[HTML]{9B9B9B} -}                    \\ \hline
Decision Tree           & Yes                     & 0.73 (0.64, 0.81)                        & 0.70 (0.61, 0.77)                        & 3                        & GAD-7                        & -                                           \\
{\color[HTML]{9B9B9B} } & {\color[HTML]{9B9B9B} } & {\color[HTML]{9B9B9B} 0.70 (0.61, 0.77)} & {\color[HTML]{9B9B9B} 0.69 (0.61, 0.77)} & {\color[HTML]{9B9B9B} 1} & {\color[HTML]{9B9B9B} GAD-7} & {\color[HTML]{9B9B9B} -}                    \\
{\color[HTML]{9B9B9B} } & {\color[HTML]{9B9B9B} } & {\color[HTML]{9B9B9B} 0.69 (0.60, 0.77)} & {\color[HTML]{9B9B9B} 0.65 (0.57, 0.73)} & {\color[HTML]{9B9B9B} 3} & {\color[HTML]{9B9B9B} SDS}   & {\color[HTML]{9B9B9B} -}                    \\ \hline
K-Nearest-Neighbors     & No                      & N.S.                                     & N.S.                                     & -                        & N.S.                         & -                                       
\end{tabular}
\vspace{-0.8mm}
\caption{\textbf{GAD-7 was the only predictor of MDD improvement across all selected models, and adding any second variable did not significantly increase AUC.} K-nearest-neighbors failed to predict significantly above chance (N.S. = not significant). 95\% confidence intervals are included in parentheses. ``Interp.'' refers to whether or not each model type is considered interpretable. Logistic regression and support vector machine coefficients are for standardized features (mean = 0, std = 1). ``Depth'' refers to maximum tree depth, and is only applicable for random forests and decision trees. A selection of alternative models that predicted significantly above chance, but were not selected by the forward process, are shown in gray: we provide results for (i) the best model with an alternative predictor (SDS) and (ii) a depth = 1 decision tree model using GAD-7, which is of special interest for interpretability. A complete set of results for the main analysis and all sensitivity analyses can be found in the supplementary material.}
\label{tab:results}
\end{table}

\begin{figure}[t] 
\centering
\includegraphics[width=0.5\columnwidth]{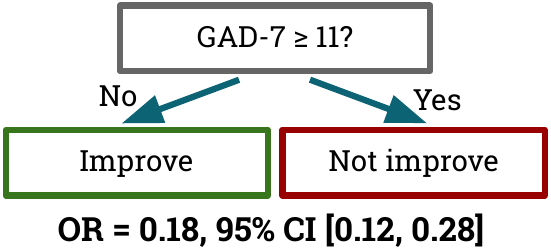}
\vspace{-2.8mm}
\caption{\textbf{A decision tree fitted to the multiply-imputed dataset predicts MDD improvement using baseline GAD-7 scores.} Participants who reported a GAD-7 score of 11 or higher were less than one-fifth as likely to experience significant MDD improvement as those with a score below 11.}
\label{fig:gad7_decision_tree}
\end{figure}

\begin{table}[]
\begin{tabular}{lll}
\textbf{Statistic}        & \textbf{Pooled Value (\%)} & \textbf{95\% Confidence Interval (\%)} \\ \hline
Sensitivity               & 73                         & 68 - 78                                \\
Specificity               & 67                         & 59 - 73                                \\
Positive Predictive Value & 76                         & 70 - 81                                \\
Negative Predictive Value & 63                         & 56 - 70                               
\end{tabular}
\vspace{-0.8mm}
\caption{\textbf{Prognostic performance of the GAD-7 $\geq$ 11 decision tree threshold for predicting MDD non-improvement.} Sensitivity is the probability that a participant who does not experience MDD improvement is correctly identified by the model (true positive rate). Specificity is the probability that a participant who does experience MDD improvement is correctly identified by the model (true negative rate). Positive Predictive Value is the probability that a participant with a GAD-7 score $\geq$ 11 truly does not experience MDD improvement. Negative Predictive Value is the probability that a participant with a GAD-7 score $<$ 11 truly does experience MDD improvement. Point estimates and confidence intervals for each of these metrics are pooled from the multiply-imputed dataset using Rubin’s rules.}
\label{tab:sens_spec_ppv_npv}
\end{table}

Logistic regression, support vector machines, random forests, and decision trees demonstrated AUC values significantly above the 0.5 chance level in predicting MDD improvement, while k-nearest-neighbors did not (Table \ref{tab:results}). All selected models used only one feature, baseline GAD-7: in no configuration did adding any additional variable increase AUC with statistical significance. Notably, the predictive performance was highly similar across these top 4 model types, with AUC and accuracy point estimates falling within narrow ranges (0.728-0.739 and 0.694-0.702 respectively) with widely overlapping confidence intervals for both. All model types other than k-nearest-neighbors also predicted significantly above the chance level using SDS as the sole predictor (indicating a negative relationship between functional disability and depression improvement), but were not selected by the forward process due to marginally lower AUC estimates. 

While all model types had similar AUC values and near-identical accuracy, decision trees are arguably the most straightforward to apply in clinical settings. The depth = 1 model provides a simple clinical heuristic, a binary threshold on a single variable, that can be applied instantly without computation. Although the depth = 3 decision tree yielded a marginally higher AUC value (0.728 vs. 0.696), the simpler and more interpretable depth = 1 tree achieved nearly identical classification accuracy (0.697 vs. 0.692 respectively; see Table \ref{tab:results}). We thus focus our primary interpretation on the depth = 1 model, as it maximizes ease of interpretability with no meaningful loss in predictive performance. 

Re-fitting a depth = 1 decision tree on the entire dataset (taking the median threshold across imputations) yields a model that classifies participants with a baseline GAD-7 score $\geq 11$ as unlikely to experience improvement, and those with GAD-7 $< 11$ as likely to experience improvement (\textbf{Fig.~\ref{fig:gad7_decision_tree}}). Notably, this threshold is similar (but not identical) to the \cite{spitzer2006brief} threshold (GAD-7 $\geq 10$) for ``moderate'' to ``severe'' generalized anxiety disorder. 

To better quantify the association of GAD-7 being 11 or higher with MDD outcomes, we calculated an odds ratio of 0.18 for improvement vs. non-improvement given GAD-7 $\geq 11$, with a 95\% confidence interval of $[0.12, 0.28]$ ($p < 0.0001$). These findings indicate that a GAD-7 score of 11 or higher reduces the odds of clinical MDD improvement by a factor of slightly less than one-fifth, with statistical significance. We used Rubin's rules to combine the odds ratio values from multiple imputations \citep{rubin1987multiple}. To explore the consistency of this finding across treatment assignment groups, we calculated the odds ratio for a GAD-7 score $\geq 11$ predicting improvement within each group separately. The association was statistically significant in all three groups: Project EVO ($0.14 \hspace{0.5em} [0.07, 0.28]$), iPST ($0.12 \hspace{0.5em} [0.04, 0.39]$), and the Health Tips active control ($0.27 \hspace{0.5em} [0.14, 0.51]$).

Viewed as a test that predicts MDD non-improvement, the proposed GAD-7 threshold has a positive predictive value of 76\% and a negative predictive value of 63\% (Table~\ref{tab:sens_spec_ppv_npv}). In other words, our results suggest that the probability that a patient with GAD-7 $<$ 11 will show clinically significant improvement in MDD is 0.63, but the probability that a patient with GAD-7 $\geq$ 11 will improve is only 0.24.

The results of both sensitivity analyses are consistent overall with those of the main analysis in that (i) GAD-7 is the single best predictor of MDD improvement, and (ii) GAD-7 $\geq 11$ is the most informative threshold for a depth = 1 decision tree, with statistically significant odds ratios. In the sensitivity analysis excluding participants with substantial missing data, the reduced statistical power meant that only a random forest model identified GAD-7 as a significant predictor during variable selection. However, an exploratory analysis of the depth = 1 decision tree in this subgroup yielded the same GAD-7 $\geq 11$ threshold with a statistically significant odds ratio. Detailed results for all sensitivity analyses are available in the supplementary material. 

While GAD-7 was the single most informative predictor of MDD improvement, SDS was also a statistically significant predictor in alternative univariate models. Yet, bivariate models combining GAD-7 and SDS did not yield significant improvements. In a post-hoc exploratory analysis, a Spearman's rank-order correlation revealed a strong, positive association between GAD-7 and SDS at baseline ($\rho = 0.67$, 95\% CI [0.61, 0.71], $p < 0.001$).

\vspace{-3mm}
\section{Discussion}
\label{sec:discuss}
\vspace{-1mm}

We predicted MDD improvement in a large cohort of participants receiving smartphone-delivered interventions. Our decision tree analysis identified a clear and clinically meaningful relationship: depressed individuals with baseline GAD-7 scores of 11 or higher were roughly one-fifth as likely to experience MDD improvement as those with lower GAD-7 scores. In terms of predictive value, there is a large difference in the probability of MDD improvement when GAD-7 $\geq$ 11 (24\%) and when GAD-7 $<$ 11 (63\%). If replicated beyond the present study, this result suggests that additional support or more intensive treatment is warranted for individuals with MDD and GAD-7 $\geq 11$ compared to people with GAD-7 $< 11$. This simple decision rule would be readily applicable in the clinic, as GAD-7 is widely administered to assess a common set of MDD comorbidities. While originally developed to screen for generalized anxiety disorder \citep{spitzer2006brief}, the GAD-7 is now well-established as a screening tool for multiple anxiety disorders \citep{kroenke2007anxiety} and as a transdiagnostic measure of global anxiety symptoms \citep{beard2014beyond}. 

Our findings highlight the utility of decision trees as a readily interpretable non-linear modeling approach. While decision trees demonstrated similar AUC scores and near-identical accuracy to logistic regression, support vector machine, and random forest models, these four model types offer different levels of interpretability - the ability to understand and explain the way the trained model makes predictions. While logistic regression and support vector machines produce coefficients that can be interpreted as the importance of each predictor, the meanings of the actual coefficient values are often unintuitive: logistic regression coefficients represent log-odds, while support vector machine coefficients define a class-separating hyperplane. Random forests are effectively uninterpretable ``black-boxes'': we cannot practicably explain how our random forests use GAD-7 to predict outcomes. In contrast, decision trees provided a GAD-7 \textit{threshold} (GAD-7 $\geq 11$) above which MDD improvement is markedly less likely. While decision trees are commonly overlooked in both modern machine learning and traditional statistical analyses, they can generate predictive rules that can be easily interpreted and applied by clinicians and thus directly inform clinical decision-making \citep{banerjee2019tree}.

In addition to the interpretability of decision trees, our machine learning-based pipeline, while computationally intensive, offers distinct advantages over traditional statistical approaches. Our framework systematically evaluated both linear models (such as logistic regression) and non-linear models (such as decision trees and random forests), allowing for the discovery of complex relationships that a single, simpler model might miss. Our methodology is also inherently robust: by combining random forest-based multiple imputation with Monte Carlo cross-validation, we rigorously accounted for uncertainty arising from both missing data and participant sampling without making strong distributional assumptions. Finally, our forward variable selection process, which included stringent criteria for model improvement and correction for multiple comparisons, provided a defense against overfitting and identified a parsimonious model while effectively handling co-linearity. While baseline GAD-7 and SDS scores were correlated, the selection algorithm identified GAD-7 as the most powerful single predictor and determined that adding SDS offered no significant improvement in predictive performance. This avoids the unstable coefficient estimates and interpretation challenges that co-linearity creates in standard multivariate regression, resulting in a more parsimonious and robust final model. 

It is important to note two limitations of this study. As would be expected from a remotely-recruited national sample for an effectiveness trial, there is a substantial proportion of missing data. We address this limitation by using a rigorous multiple imputation approach, and with a sensitivity analysis that reduces the proportion of missing data by excluding participants with only minimal baseline data. Furthermore, in a supplementary complete-case analysis, we showed that GAD-7 predicts MDD improvement in a logistic regression model, confirming a statistically significant association in the raw, unimputed data (see Supplementary Section 3 for details). While this simpler analysis corroborates our findings, our primary machine learning-based methodology offers several distinct advantages. First, our analysis mitigates the risk of attrition bias, a significant concern given that the complete-case analysis excludes the majority of study participants. Second, multiple imputation increases statistical power while accounting for uncertainty introduced by missing data. Third, our approach is hypothesis-free, identifying GAD-7 as the best predictor of MDD outcomes without a priori assumptions. Finally, our analysis yielded a clinically intuitive threshold (GAD-7 $\geq$ 11) rather than a less interpretable odds ratio for an ordinal predictor. 

A second limitation is that our analysis cannot establish causal associations. One possible explanation for the association between baseline GAD-7 and MDD improvement is that anxiety hinders participants' engagement with smartphone-delivered interventions. Given findings from previous studies that comorbid anxiety reduces pharmacological treatment response in MDD (e.g., \citet{fava2008difference, saveanu2015international, dold2017clinical}), it is also conceivable that comorbid anxiety hinders treatment response through a mechanism that is independent of treatment modality. A third possibility, which our results support most strongly, is that anxiety is a prognostic factor for poorer short-term MDD outcomes, an effect that is independent of both treatment assignment and treatment adherence. The original clinical trial analysis found higher MDD remission rates for moderately depressed participants in the Project EVO and iPST groups compared to the Health Tips control group, despite also reporting that the majority of participants in the active treatment groups did not download the assigned intervention app \citep{arean2016use}.  Our analysis, which used different inclusion criteria and was designed to identify outcome predictors with corrections for multiple comparisons, not treatment effectiveness specifically, did not identify treatment assignment as a significant predictor. This is consistent with our finding that GAD-7 $\geq 11$ predicted lower odds of improvement within both the active treatment groups and the Health Tips control group, supporting the interpretation that baseline anxiety is a general prognostic factor for poorer short-term MDD outcomes in this context, rather than a predictor of response to a specific treatment. While the point estimates for the odds ratios are consistent with a stronger association between baseline anxiety and MDD improvement in the active treatment groups than in the control group, their wide and overlapping confidence intervals preclude a definitive conclusion about an interaction between baseline anxiety and treatment assignment. Ultimately, our study cannot distinguish between treatment-related, treatment-unrelated, or combined mechanisms for the observed association between baseline GAD-7 scores and PHQ-9 score reductions. 

The link between higher baseline anxiety and lower odds of MDD improvement in this setting may have clinically actionable implications. For example, if this association is partly due to patients with comorbid anxiety struggling to engage with smartphone-delivered interventions, these patients might need additional support or different therapeutic approaches. Regardless of the underlying mechanisms, one practical implication of our results is that future randomized controlled trials of smartphone-delivered interventions for MDD should consider stratifying by baseline anxiety levels. 

\vspace{-3mm}
\subsubsection*{Acknowledgements}

Morgan B. Talbot's time was partially supported by the National Institute of General Medical Sciences under Award T32GM144273, and partially supported by Massachusetts Institute of Technology through the David and Beatrice Yamron Fellowship. Dr. Lipschitz’s time was partially supported by the National Institute of Mental Health (NIMH) under Grant MH120324. Dr. Costilla-Reyes' time was partially supported by the National Science Foundation (NSF) under Award 1918839. The content of this paper is solely the responsibility of the authors and does not necessarily represent the official views of any of the above organizations.

\bibliographystyle{elsarticle-harv} 
\vspace{-3mm}
\bibliography{references.bib}

\begin{thebibliography}{24}
\expandafter\ifx\csname natexlab\endcsname\relax\def\natexlab#1{#1}\fi
\providecommand{\url}[1]{\texttt{#1}}
\providecommand{\href}[2]{#2}
\providecommand{\path}[1]{#1}
\providecommand{\DOIprefix}{doi:}
\providecommand{\ArXivprefix}{arXiv:}
\providecommand{\URLprefix}{URL: }
\providecommand{\Pubmedprefix}{pmid:}
\providecommand{\doi}[1]{\href{http://dx.doi.org/#1}{\path{#1}}}
\providecommand{\Pubmed}[1]{\href{pmid:#1}{\path{#1}}}
\providecommand{\bibinfo}[2]{#2}
\ifx\xfnm\relax \def\xfnm[#1]{\unskip,\space#1}\fi
\bibitem[{Anguera et~al.(2016)Anguera, Jordan, Castaneda, Gazzaley and
  Are{\'a}n}]{anguera2016conducting}
\bibinfo{author}{Anguera, J.A.}, \bibinfo{author}{Jordan, J.T.},
  \bibinfo{author}{Castaneda, D.}, \bibinfo{author}{Gazzaley, A.},
  \bibinfo{author}{Are{\'a}n, P.A.}, \bibinfo{year}{2016}.
\newblock \bibinfo{title}{Conducting a fully mobile and randomised clinical
  trial for depression: access, engagement and expense}.
\newblock \bibinfo{journal}{BMJ Innovations} \bibinfo{volume}{2}.
\bibitem[{Arean et~al.(2016)Arean, Hallgren, Jordan, Gazzaley, Atkins, Heagerty
  and Anguera}]{arean2016use}
\bibinfo{author}{Arean, P.A.}, \bibinfo{author}{Hallgren, K.A.},
  \bibinfo{author}{Jordan, J.T.}, \bibinfo{author}{Gazzaley, A.},
  \bibinfo{author}{Atkins, D.C.}, \bibinfo{author}{Heagerty, P.J.},
  \bibinfo{author}{Anguera, J.A.}, \bibinfo{year}{2016}.
\newblock \bibinfo{title}{The use and effectiveness of mobile apps for
  depression: results from a fully remote clinical trial}.
\newblock \bibinfo{journal}{Journal of Medical Internet Research}
  \bibinfo{volume}{18}, \bibinfo{pages}{e330}.
\bibitem[{Banerjee et~al.(2019)Banerjee, Reynolds, Andersson and
  Nallamothu}]{banerjee2019tree}
\bibinfo{author}{Banerjee, M.}, \bibinfo{author}{Reynolds, E.},
  \bibinfo{author}{Andersson, H.B.}, \bibinfo{author}{Nallamothu, B.K.},
  \bibinfo{year}{2019}.
\newblock \bibinfo{title}{Tree-based analysis: a practical approach to create
  clinical decision-making tools}.
\newblock \bibinfo{journal}{Circulation: Cardiovascular Quality and Outcomes}
  \bibinfo{volume}{12}, \bibinfo{pages}{e004879}.
\bibitem[{Beard and Bj{\"o}rgvinsson(2014)}]{beard2014beyond}
\bibinfo{author}{Beard, C.}, \bibinfo{author}{Bj{\"o}rgvinsson, T.},
  \bibinfo{year}{2014}.
\newblock \bibinfo{title}{Beyond generalized anxiety disorder: psychometric
  properties of the {GAD-7} in a heterogeneous psychiatric sample}.
\newblock \bibinfo{journal}{Journal of anxiety disorders} \bibinfo{volume}{28},
  \bibinfo{pages}{547--552}.
\bibitem[{Benjamini and Hochberg(1995)}]{benjamini1995controlling}
\bibinfo{author}{Benjamini, Y.}, \bibinfo{author}{Hochberg, Y.},
  \bibinfo{year}{1995}.
\newblock \bibinfo{title}{Controlling the false discovery rate: a practical and
  powerful approach to multiple testing}.
\newblock \bibinfo{journal}{Journal of the Royal Statistical Society: Series B
  (Methodological)} \bibinfo{volume}{57}, \bibinfo{pages}{289--300}.
\bibitem[{Bush et~al.(1998)Bush, Kivlahan, McDonell, Fihn, Bradley and
  Project}]{bush1998audit}
\bibinfo{author}{Bush, K.}, \bibinfo{author}{Kivlahan, D.R.},
  \bibinfo{author}{McDonell, M.B.}, \bibinfo{author}{Fihn, S.D.},
  \bibinfo{author}{Bradley, K.A.}, \bibinfo{author}{Project, A.C.Q.I.},
  \bibinfo{year}{1998}.
\newblock \bibinfo{title}{The {AUDIT} alcohol consumption questions
  ({AUDIT-C}): an effective brief screening test for problem drinking}.
\newblock \bibinfo{journal}{Archives of Internal Medicine}
  \bibinfo{volume}{158}, \bibinfo{pages}{1789--1795}.
\bibitem[{Dold et~al.(2017)Dold, Bartova, Souery, Mendlewicz, Serretti,
  Porcelli, Zohar, Montgomery and Kasper}]{dold2017clinical}
\bibinfo{author}{Dold, M.}, \bibinfo{author}{Bartova, L.},
  \bibinfo{author}{Souery, D.}, \bibinfo{author}{Mendlewicz, J.},
  \bibinfo{author}{Serretti, A.}, \bibinfo{author}{Porcelli, S.},
  \bibinfo{author}{Zohar, J.}, \bibinfo{author}{Montgomery, S.},
  \bibinfo{author}{Kasper, S.}, \bibinfo{year}{2017}.
\newblock \bibinfo{title}{Clinical characteristics and treatment outcomes of
  patients with major depressive disorder and comorbid anxiety
  disorders-results from a {E}uropean multicenter study}.
\newblock \bibinfo{journal}{Journal of Psychiatric Research}
  \bibinfo{volume}{91}, \bibinfo{pages}{1--13}.
\bibitem[{Fava et~al.(2008)Fava, Rush, Alpert, Balasubramani, Wisniewski,
  Carmin, Biggs, Zisook, Leuchter, Howland et~al.}]{fava2008difference}
\bibinfo{author}{Fava, M.}, \bibinfo{author}{Rush, A.J.},
  \bibinfo{author}{Alpert, J.E.}, \bibinfo{author}{Balasubramani, G.},
  \bibinfo{author}{Wisniewski, S.R.}, \bibinfo{author}{Carmin, C.N.},
  \bibinfo{author}{Biggs, M.M.}, \bibinfo{author}{Zisook, S.},
  \bibinfo{author}{Leuchter, A.}, \bibinfo{author}{Howland, R.}, et~al.,
  \bibinfo{year}{2008}.
\newblock \bibinfo{title}{Difference in treatment outcome in outpatients with
  anxious versus nonanxious depression: a {STAR*D} report}.
\newblock \bibinfo{journal}{American Journal of Psychiatry}
  \bibinfo{volume}{165}, \bibinfo{pages}{342--351}.
\bibitem[{Gaynes et~al.(2020)Gaynes, Lux, Gartlehner, Asher, Forman-Hoffman,
  Green, Boland, Weber, Randolph, Bann et~al.}]{gaynes2020defining}
\bibinfo{author}{Gaynes, B.N.}, \bibinfo{author}{Lux, L.},
  \bibinfo{author}{Gartlehner, G.}, \bibinfo{author}{Asher, G.},
  \bibinfo{author}{Forman-Hoffman, V.}, \bibinfo{author}{Green, J.},
  \bibinfo{author}{Boland, E.}, \bibinfo{author}{Weber, R.P.},
  \bibinfo{author}{Randolph, C.}, \bibinfo{author}{Bann, C.}, et~al.,
  \bibinfo{year}{2020}.
\newblock \bibinfo{title}{Defining treatment-resistant depression}.
\newblock \bibinfo{journal}{Depression and Anxiety} \bibinfo{volume}{37},
  \bibinfo{pages}{134--145}.
\bibitem[{Greenberg et~al.(2024)Greenberg, Weingarden, Hoeppner,
  Berger-Gutierrez, Klare, Snorrason, Costilla-Reyes, Talbot, Daniel,
  Vanderkruik et~al.}]{greenberg2024predicting}
\bibinfo{author}{Greenberg, J.L.}, \bibinfo{author}{Weingarden, H.},
  \bibinfo{author}{Hoeppner, S.S.}, \bibinfo{author}{Berger-Gutierrez, R.M.},
  \bibinfo{author}{Klare, D.}, \bibinfo{author}{Snorrason, I.},
  \bibinfo{author}{Costilla-Reyes, O.}, \bibinfo{author}{Talbot, M.},
  \bibinfo{author}{Daniel, K.E.}, \bibinfo{author}{Vanderkruik, R.C.}, et~al.,
  \bibinfo{year}{2024}.
\newblock \bibinfo{title}{Predicting response to a smartphone-based
  cognitive-behavioral therapy for body dysmorphic disorder}.
\newblock \bibinfo{journal}{Journal of Affective Disorders}
  \bibinfo{volume}{355}, \bibinfo{pages}{106--114}.
\bibitem[{Kroenke et~al.(2001)Kroenke, Spitzer and Williams}]{kroenke2001phq}
\bibinfo{author}{Kroenke, K.}, \bibinfo{author}{Spitzer, R.L.},
  \bibinfo{author}{Williams, J.B.}, \bibinfo{year}{2001}.
\newblock \bibinfo{title}{The {PHQ-9}: validity of a brief depression severity
  measure}.
\newblock \bibinfo{journal}{Journal of General Internal Medicine}
  \bibinfo{volume}{16}, \bibinfo{pages}{606--613}.
\bibitem[{Kroenke et~al.(2007)Kroenke, Spitzer, Williams, Monahan and
  L{\"o}we}]{kroenke2007anxiety}
\bibinfo{author}{Kroenke, K.}, \bibinfo{author}{Spitzer, R.L.},
  \bibinfo{author}{Williams, J.B.}, \bibinfo{author}{Monahan, P.O.},
  \bibinfo{author}{L{\"o}we, B.}, \bibinfo{year}{2007}.
\newblock \bibinfo{title}{Anxiety disorders in primary care: prevalence,
  impairment, comorbidity, and detection}.
\newblock \bibinfo{journal}{Annals of internal medicine} \bibinfo{volume}{146},
  \bibinfo{pages}{317--325}.
\bibitem[{Linardon et~al.(2024)Linardon, Torous, Firth, Cuijpers, Messer and
  Fuller-Tyszkiewicz}]{linardon2024current}
\bibinfo{author}{Linardon, J.}, \bibinfo{author}{Torous, J.},
  \bibinfo{author}{Firth, J.}, \bibinfo{author}{Cuijpers, P.},
  \bibinfo{author}{Messer, M.}, \bibinfo{author}{Fuller-Tyszkiewicz, M.},
  \bibinfo{year}{2024}.
\newblock \bibinfo{title}{Current evidence on the efficacy of mental health
  smartphone apps for symptoms of depression and anxiety. a meta-analysis of
  176 randomized controlled trials}.
\newblock \bibinfo{journal}{World Psychiatry} \bibinfo{volume}{23},
  \bibinfo{pages}{139--149}.
\bibitem[{Moreno-Agostino et~al.(2021)Moreno-Agostino, Wu, Daskalopoulou,
  Hasan, Huisman and Prina}]{moreno2021global}
\bibinfo{author}{Moreno-Agostino, D.}, \bibinfo{author}{Wu, Y.T.},
  \bibinfo{author}{Daskalopoulou, C.}, \bibinfo{author}{Hasan, M.T.},
  \bibinfo{author}{Huisman, M.}, \bibinfo{author}{Prina, M.},
  \bibinfo{year}{2021}.
\newblock \bibinfo{title}{Global trends in the prevalence and incidence of
  depression: a systematic review and meta-analysis}.
\newblock \bibinfo{journal}{Journal of Affective Disorders}
  \bibinfo{volume}{281}, \bibinfo{pages}{235--243}.
\bibitem[{Pedregosa et~al.(2011)Pedregosa, Varoquaux, Gramfort, Michel,
  Thirion, Grisel, Blondel, Prettenhofer, Weiss, Dubourg
  et~al.}]{pedregosa2011scikit}
\bibinfo{author}{Pedregosa, F.}, \bibinfo{author}{Varoquaux, G.},
  \bibinfo{author}{Gramfort, A.}, \bibinfo{author}{Michel, V.},
  \bibinfo{author}{Thirion, B.}, \bibinfo{author}{Grisel, O.},
  \bibinfo{author}{Blondel, M.}, \bibinfo{author}{Prettenhofer, P.},
  \bibinfo{author}{Weiss, R.}, \bibinfo{author}{Dubourg, V.}, et~al.,
  \bibinfo{year}{2011}.
\newblock \bibinfo{title}{Scikit-learn: {M}achine learning in {P}ython}.
\newblock \bibinfo{journal}{Journal of Machine Learning Research}
  \bibinfo{volume}{12}, \bibinfo{pages}{2825--2830}.
\bibitem[{Rubin(1987)}]{rubin1987multiple}
\bibinfo{author}{Rubin, D.B.}, \bibinfo{year}{1987}.
\newblock \bibinfo{title}{Multiple imputation for nonresponse in surveys}.
\newblock \bibinfo{publisher}{John Wiley \& Sons}, \bibinfo{address}{New York}.
\bibitem[{{Sage Bionetworks}(2018)}]{BrightenStudyPortal}
\bibinfo{author}{{Sage Bionetworks}}, \bibinfo{year}{2018}.
\newblock \bibinfo{title}{Brighten study public researcher portal}.
\newblock
  \bibinfo{howpublished}{\url{https://www.synapse.org/Synapse:syn10848316}}.
\newblock \bibinfo{note}{Accessed: 2025-07-22, Synapse ID: syn10848316}.
\bibitem[{Saveanu et~al.(2015)Saveanu, Etkin, Duchemin, Goldstein-Piekarski,
  Gyurak, Debattista, Schatzberg, Sood, Day, Palmer
  et~al.}]{saveanu2015international}
\bibinfo{author}{Saveanu, R.}, \bibinfo{author}{Etkin, A.},
  \bibinfo{author}{Duchemin, A.M.}, \bibinfo{author}{Goldstein-Piekarski, A.},
  \bibinfo{author}{Gyurak, A.}, \bibinfo{author}{Debattista, C.},
  \bibinfo{author}{Schatzberg, A.F.}, \bibinfo{author}{Sood, S.},
  \bibinfo{author}{Day, C.V.}, \bibinfo{author}{Palmer, D.M.}, et~al.,
  \bibinfo{year}{2015}.
\newblock \bibinfo{title}{The international study to predict optimized
  treatment in depression ({iSPOT-D}): outcomes from the acute phase of
  antidepressant treatment}.
\newblock \bibinfo{journal}{Journal of Psychiatric Research}
  \bibinfo{volume}{61}, \bibinfo{pages}{1--12}.
\bibitem[{Sextl-Pl{\"o}tz et~al.(2024)Sextl-Pl{\"o}tz, Steinhoff, Baumeister,
  Cuijpers, Ebert and Zarski}]{sextl2024systematic}
\bibinfo{author}{Sextl-Pl{\"o}tz, T.}, \bibinfo{author}{Steinhoff, M.},
  \bibinfo{author}{Baumeister, H.}, \bibinfo{author}{Cuijpers, P.},
  \bibinfo{author}{Ebert, D.D.}, \bibinfo{author}{Zarski, A.C.},
  \bibinfo{year}{2024}.
\newblock \bibinfo{title}{A systematic review of predictors and moderators of
  treatment outcomes in internet-and mobile-based interventions for
  depression}.
\newblock \bibinfo{journal}{Internet Interventions} , \bibinfo{pages}{100760}.
\bibitem[{Sheehan(1983)}]{sheehan1983anxiety}
\bibinfo{author}{Sheehan, D.}, \bibinfo{year}{1983}.
\newblock \bibinfo{title}{The Anxiety Disease}.
\newblock \bibinfo{publisher}{Scribner}, \bibinfo{address}{New York}.
\bibitem[{Spitzer et~al.(2006)Spitzer, Kroenke, Williams and
  L{\"o}we}]{spitzer2006brief}
\bibinfo{author}{Spitzer, R.L.}, \bibinfo{author}{Kroenke, K.},
  \bibinfo{author}{Williams, J.B.}, \bibinfo{author}{L{\"o}we, B.},
  \bibinfo{year}{2006}.
\newblock \bibinfo{title}{A brief measure for assessing generalized anxiety
  disorder: the {GAD-7}}.
\newblock \bibinfo{journal}{Archives of Internal Medicine}
  \bibinfo{volume}{166}, \bibinfo{pages}{1092--1097}.
\bibitem[{Thornicroft et~al.(2017)Thornicroft, Chatterji, Evans-Lacko, Gruber,
  Sampson, Aguilar-Gaxiola, Al-Hamzawi, Alonso, Andrade, Borges
  et~al.}]{thornicroft2017undertreatment}
\bibinfo{author}{Thornicroft, G.}, \bibinfo{author}{Chatterji, S.},
  \bibinfo{author}{Evans-Lacko, S.}, \bibinfo{author}{Gruber, M.},
  \bibinfo{author}{Sampson, N.}, \bibinfo{author}{Aguilar-Gaxiola, S.},
  \bibinfo{author}{Al-Hamzawi, A.}, \bibinfo{author}{Alonso, J.},
  \bibinfo{author}{Andrade, L.}, \bibinfo{author}{Borges, G.}, et~al.,
  \bibinfo{year}{2017}.
\newblock \bibinfo{title}{Undertreatment of people with major depressive
  disorder in 21 countries}.
\newblock \bibinfo{journal}{British Journal of Psychiatry}
  \bibinfo{volume}{210}, \bibinfo{pages}{119--124}.
\bibitem[{Un{\"u}tzer et~al.(2002)Un{\"u}tzer, Katon, Callahan, Williams~Jr,
  Hunkeler, Harpole, Hoffing, Della~Penna, No{\"e}l, Lin
  et~al.}]{unutzer2002collaborative}
\bibinfo{author}{Un{\"u}tzer, J.}, \bibinfo{author}{Katon, W.},
  \bibinfo{author}{Callahan, C.M.}, \bibinfo{author}{Williams~Jr, J.W.},
  \bibinfo{author}{Hunkeler, E.}, \bibinfo{author}{Harpole, L.},
  \bibinfo{author}{Hoffing, M.}, \bibinfo{author}{Della~Penna, R.D.},
  \bibinfo{author}{No{\"e}l, P.H.}, \bibinfo{author}{Lin, E.H.}, et~al.,
  \bibinfo{year}{2002}.
\newblock \bibinfo{title}{Collaborative care management of late-life depression
  in the primary care setting: a randomized controlled trial}.
\newblock \bibinfo{journal}{Journal of the American Medical Association (JAMA)}
  \bibinfo{volume}{288}, \bibinfo{pages}{2836--2845}.
\bibitem[{Wilson(2020)}]{wilson2020miceranger}
\bibinfo{author}{Wilson, S.}, \bibinfo{year}{2020}.
\newblock \bibinfo{title}{{miceRanger}: {M}ultiple imputation by chained
  equations with random forests}.
\newblock
  \bibinfo{howpublished}{\url{https://cran.r-project.org/web/packages/miceRanger/index.html}}.
\newblock \bibinfo{note}{CRAN package (R)}.

\end{thebibliography}

\ifarXiv
    \includepdf[pages=-]{\supplementfilename}
\fi

\end{document}